\def\ref{\par \noindent 
\hangindent=1pc
\hangafter=-1}

\def\etal{{\it et al.\thinspace}}

\def\approxlt{\mathrel{\hbox{\rlap{\lower .5ex \hbox {$\sim$}}
	\raise .15 ex \hbox{$<$}}}}
\def\approxgt{\mathrel{\hbox{\rlap{\lower .5ex \hbox {$\sim$}}
	\raise .15 ex \hbox{$>$}}}}
\mathchardef\twiddle="2218

\def\multleft#1{\hbox to size{\vbox {\halign {\lft{##}\cr #1}}\hfill}\par}
\def\multright#1{\hbox to size{\vbox {\halign {\rt{##}\cr #1}}\hfill}\par}

\def\<{\thinspace}

\def\spose#1{\hbox to 0pt{#1\hss}}
\def\simlt{\mathrel{\spose{\lower 3pt\hbox{$\mathchar"218$}}
     \raise 2.0pt\hbox{$\mathchar"13C$}}}
\def\simgt{\mathrel{\spose{\lower 3pt\hbox{$\mathchar"218$}}
     \raise 2.0pt\hbox{$\mathchar"13E$}}}
\def\lsim{\rlap{$<$}{\lower 1.0ex\hbox{$\sim$}}}
\def\gsim{\rlap{$>$}{\lower 1.0ex\hbox{$\sim$}}}

\def\ref{\par \noindent \hangindent=3pc \hangafter=1}

\def\titlep{\footline={\ifnum\pageno=1 \hss\else\hss\tenrm\folio
\hss\fi}}

\def\double_under#1{\underline{\underline{#1}}}

\def\K{{\rm\thinspace K}}

\def\Msun{\hbox{$\rm\thinspace M_{\odot}$}}

\documentstyle{article}

\begin{document}

\title{`FIRST LIGHT' IN THE UNIVERSE;\\
WHAT ENDED THE ``DARK AGE''?}

\author{MARTIN J REES\\\\
Institute of Astronomy,
Madingley Road\\ Cambridge, CB3 OHA\\
United Kingdom\\
Email: mjr@ast.cam.ac.uk}
\date{\phantom{-}}

\maketitle

\begin{abstract}
  The universe would have been completely dark between the epoch of
recombination and the development of the first non-linear
structure. But at redshifts beyond 5 -- perhaps even beyond 20 --
stars formed within `subgalaxies' and created the first heavy
elements; these same systems (together perhaps with `miniquasars')
generated the UV radiation that ionized the IGM, and maybe also the
first significant magnetic fields.  Although we can already probe back
to $z \simeq 5$, these very first objects may be so faint that their
detection must await next-generation optical and infrared
telescopes. Observations in other wavebands may offer indirect clues
to when reionization occurred. Despite the rapid improvements in
numerical simulations, the processes of star formation and feedback
are likely to remain a challenge for the next decade.
\end{abstract}

\section{INTRODUCTION}

One of the outstanding achievements of cosmology is that the state of
the universe when it was only a few seconds old seems to be well
understood. The details have firmed up, and we can make confident
predictions about primordial neutrinos, and He and D
nucleosynthesis. This progress, spanning the last 30 years, owed a
lot, on the theoretical side, to David Schramm and his Chicago
colleagues. The way the universe cools, and eventually recombines, and
the evolution of the (linear) perturbations that imprint angular
structures on the microwave background, is also well understood. But
this gratifying simplicity ends when primordial imhomogeneities and
density contrasts evolve into the non-linear regime.

    The Universe literally entered a dark age about 300,000 years
after the big bang, when the primordial radiation cooled below 3000K
and shifted into the infrared. Unless there were some photon input from (for instance) decaying particles, or string loops, darkness would have persisted until the first non-linearities developed
into gravitationally-bound systems, whose internal evolution gave rise
to stars, or perhaps to more massive bright objects.

   Spectroscopy from the new generation of 8-10 metre telescopes now
complements the sharp imaging of the Hubble Space Telescope (HST);
these instruments are together elucidating the history of star
formation, galaxies and clustering back, at least, to redshifts $z =
5$. Our knowledge of these eras is no longer restricted to
`pathological' objects such as extreme AGNs -- this is one of the
outstanding astronomical advances of recent years.  In addition,
quasar spectra (the Lyman forest, etc) are now observable with much
improved resolution and signal-to-noise; they offer probes of the
clumping, temperature, and composition of diffuse gas on galactic (and
smaller) scales over an equally large redshift range, rather as ice
cores enable geophysicists to probe climatic history.

    Detailed sky maps of the microwave background (CMB) temperature
(and perhaps its polarization as well) will soon offer direct
diagnostics of the initial fluctuations from which the present-day
large-scale structure developed. Most of the photons in this
background have travelled uninterruptedly since the recombination
epoch at $z = 1000$, when the fluctuations were still in the linear
regime.  We may also, in the next few years, discover the nature of
the dark matter; computer simulations of structure formation will not
only include gravity, but will incorporate the gas dynamics and
radiation of the baryonic component in a sophisticated way.

      But these advances may still leave us, several years from now,
uncertain about the quantitative details of the whole era from $10^6$ to
$10^9$ years -- the formation of the first stars, the first supernovae,
the first heavy elements; and how and when the intergalactic medium
was reionized.  Even by the time Planck/Surveyor and the Next
Generation Space Telescope (NGST) have been launched, we may still be
unable to compute crucial things like the star formation efficiency,
feedback from supernovae. etc -- processes that `semi-analytic' models
for galactic evolution now parametrise in a rather ad hoc way.
    
And CMB fluctuations will still be undiscernable on the very small
angular scales that correspond to subgalactic structures, which, in
any hierarchical (`bottom up') scenario would be the first
non-linearities to develop. So the `dark age' is likely to remain a
topic for lively controversy at least for the next decade.
   
\section{COSMOGONIC PRELIMINARIES:\\ MOLECULAR HYDROGEN AND UV\\ FEEDBACK}

\subsection{The H{\boldmath $_2$} cooling regime} 

     Detailed studies of structure formation generally focus on some
variant of the cold dark matter (CDM) cosmogony -- with a specific
choice for $\Omega_{\rm CDM}$, $\Omega_b$ and $\Lambda$. Even if such
a model turns out to be oversimplified, it offers a useful 'template'
whose main features apply generically to any `bottom up' model for
structure formation.  There is no minimum scale for gravitational
aggregation of the CDM. However, the baryonic gas does not `feel' the
very smallest clumps, which have very small binding energies: pressure
opposes condensation of the gas on scales below a (time dependent)
Jeans scale -- roughly, the size of a comoving sphere whose boundary
expands at the sound speed.

The overdense clumps of CDM within which `first light' occurs must
provide a deep enough potential well to pull the gas into them . But
they must also -- a somewhat more stringent requirement -- yield,
after virialisation, a gas temperature such that radiative cooling is
efficient enough to allow the gas to contract further.  The dominant
coolant for gas of primordial composition is molecular hydrogen. This
has been considered by many authors, from the 1960s onwards; see
recent discussions by, for instance, Tegmark \etal (1997), Haiman,
Rees and Loeb (1997),  Haiman, Abel  and
Rees (1999).  In a uniformly expanding universe, only about $10^{-6}$
of the post-recombination hydrogen is in the form of H$_2$.  However this
rises to $10^{-4}$ within collapsing regions -- high enough to permit
cooling at temperatures above a few hundred degrees.

         So the first `action' would have occurred within clumps with
virial temperatures of a few hundred degrees (corresponding to a
virial velocity of 2-3 km/s). Their total mass is of order $10^5 \Msun$ ;
the baryonic mass is smaller by a factor $\Omega_b / \Omega_{\rm CDM}$.  

The gas falling into such a clump exhibits filamentary substructure:
the contraction is almost isothermal, so the Jeans mass decreases as
the density rises. Abel, Bryan and Norman (1999) have simulated the
collapse, taking account of radiative transfer in the molecular lines,
up to $10^{12}$ times the turnaround density; by that stage the Jeans
mass (and the size of the smallest bound subclumps) has dropped to
$50-100 \Msun$.

There is still a large gap to be bridged between the endpoint of these
impressive simulations and the formation of `protostars'.
Fragmentation could continue down to smaller masses; on the other
hand, there could be no further fragmentation -- indeed, as Bromm, Coppi 
 and Larson (1999) argue, infall onto the largest blobs could lead to masses much
higher than $100 \Msun$.

And when even one star has formed, further uncertainties
ensue. Radiation or winds may expel uncondensed material from the
shallow potential wells, and exert the kind of feedback familiar from
studies of giant molecular clouds in our own Galaxy.  In addition to
this local feedback, there is a non-local effect due to UV radiation.
Photons of $h\nu > 11.18$ eV can photodissociate H$_2$, as first
calculated by Stecher and Williams (1967). These photons, softer than
the Lyman limit, can penetrate a high column density of HI and destroy
molecules in virialised and collapsing clouds. H$_2$ cooling would be
quenched if there were a UV background able to dissociate the
molecules as fast as they form. The effects within clouds have been
calculated by Haiman, Abel and Rees (1999) and Ciardi, Ferrara and
Abel (1999).

(If the radiation from the first objects had a non-thermal component
extending up to KeV energies, as it might if a contribution came from
accreting compact objects or supernovae, then there is a
counterbalancing positive feedback.  X-ray photons penetrate HI,
producing photoelectrons (which themselves cause further collisional ionizationwhile being slowed down and thermalised); these electrons
then catalyse further H$_2$ formation via H$^-$).
         
   It seems most likely that the negative feedback due to
photoionization is dominant. When the UV background gets above a
certain threshold, H$_2$ is prevented from forming and molecular
cooling is suppressed.  Under all plausible assumptions about UV
spectral shape, etc, this threshold is reached well before there has
been enough UV production to ionize most of the medium.  Therefore,
only a small fraction of the UV that ionized the IGM can have been
produced in systems where star formation was triggered by molecular
cooling.

\subsection{The atomic-cooling stage}

 An atomic H-He mixture behaves adiabatically unless T is as high as
8-10 thousand degrees, when excitation of Lyman alpha by the
Maxwellian tail of the electrons provides efficient cooling whose rate
rises steeply with temperature.

     When H$_2$ cooling has been quenched, primordial gas cannot
therefore cool and fragment within bound systems unless their virial
temperature reaches $10^4 \, \K$. The corresponding mass is $\sim 10^8 \Msun$.
Most of the UV that ionized the IGM therefore came from stars (or
perhaps from accreting black holes) that formed within systems of
total mass $ \simgt 10^8 \Msun$.
   
\section{THE EPOCH OF IONIZATION\\ BREAKTHROUGH}

\subsection{UV production in `subgalaxies'}

   The IGM would have remained predominantly neutral until
`subgalaxies', with total (dark matter) masses above $10^8 \Msun$ and
virial velocities 20 km/s, had generated enough photoionizing flux
from O-B stars, or perhaps accreting black holes (see Loeb (1999) and
references cited therein).
       
 How many of these `subgalaxies' formed, and how bright each one would
be, depends on another big uncertainty: the IMF and formation
efficiency for the Population III objects.

  The gravitational aspects of clustering can all be modeled
convincingly by computer simulations. So also, now, can the dynamics
of the baryonic (gaseous) component -- including shocks and radiative
cooling.  The huge dynamic range of the star-formation process cannot
be tracked computationally up to the densities at which individual
stars condense out.  But the nature of the simulation changes as soon
as the first stars (or other compact objects) form.  The first stars
(or other compact objects) exert crucial feedback -- the remaining gas
is heated by ionizing radiation, and perhaps also by an injection of
kinetic energy via winds and even supernova explosions -- which is
even harder to model, being sensitive to the IMF, and to further
uncertain physics.

     Three major uncertainties  are:

(i) What is the IMF of the first stellar population? The high-mass
stars are the ones that provide efficient (and relatively prompt)
feedback. It plainly makes a big difference whether these are the
dominant type of stars, or whether the initial IMF rises steeply
towards low masses (or is bimodal), so that very many faint stars form
before there is a significant feedback.  The Population III objects
form in an unmagnetised medium of pure H and He, bathed in background
radiation that may be hotter than 50 K when the action starts (at
redshift $z$ the ambient temperature is of course $2.7(1 + z)$ K). Would
these conditions favour a flatter or a steeper IMF than we observed
today?  This is completely unclear: the density may become so high
that fragmentation proceeds to very low masses (despite the higher
temperature and absence of coolants other than molecular hydrogen); on
the other hand, massive stars may be more favoured than at the present
epoch. Indeed, fragmentation could even be so completely inhibited
that the first things to form are supermassive holes.

(ii) Quite apart from the uncertainty in the IMF, it is also unclear
what fraction of the baryons that fall into a clump would actually be
incorporated into stars before being re-ejected. The retained fraction
would almost certainly be an increasing function of virial velocity:
gas more readily escapes from shallow potential wells.
 
(iii) The influence of the Population III objects depends on how much
of their radiation escapes into the IGM. Much of the Lyman continuum
emitted within a `subgalaxy' could, for instance, be absorbed within
it. The total number of massive stars or accreting holes needed to
build up the UV background shortward of the Lyman limit and ionize the
IGM, and the concomitant contamination by heavy elements, would then
be greater.

  All these three uncertainties would, for a given fluctuation
spectrum, affect the redshift at which molecules were destroyed, and
at which full ionization occurred. Perhaps I'm being pessimistic, but
I doubt that either observations or theoretical progress will have
eliminated these uncertainties about the `dark age' even by the time
NGST flies.

\subsection{How uncertain is the ionization epoch?}

   Even if we knew exactly what the initial fluctuations were, and
when the first bound systems on each scale formed, the above-mentioned
uncertainties would render the ionization redshift is uncertain by at
least a factor of 2.  This can be easily seen as follows:

      Ionization breakthrough requires at least 1 photon for each
ionized baryon in the IGM (one photon per baryon is obviously needed;
extra photons are needed to balance recombinations, which are more
important in clumps and filaments than in underdense regions).  An OB
star produces $10^4 - 10^5$ ionizing photons for each constituent
baryon, so (again in very round numbers) $10^{-3}$ of the baryons must
condense into stars with a standard IMF to supply the requisite UV.
Photoionization will be discussed in Madau's contribution to this
conference. Earlier references include Ciardi and Ferrara (1997),
Gnedin and Ostriker (1998), Madau, Haardt and Rees (1999) and Gnedin
(1999).

     We can then contrast two cases:

     (A)  If the star formation were efficient, in the sense that all
the baryons that `went non-linear', and fell into a CDM clump larger
than the Jeans mass, turned into stars, then the rare 3-$\sigma$ peaks on
mass-scales $10^8 \Msun$ would suffice.

  On the other hand:

(B) Star formation could plausibly be so inefficient that less than 1
percent of the baryons in a pregalaxy condense into stars, the others
being expelled by stellar winds, supernovae, etc., In this case,
production of the necessary UV would have to await the collapse of
more typical peaks (1.5-$\sigma$, for instance).

      A 1.5-$\sigma$ peak has an initial amplitude only half that of a 
3-$\sigma$ peak, and would therefore collapse at a value of $(1 + z)$ that was
lower by a factor of 2.  For plausible values of the fluctuation
amplitude this could change $z_i$ from 15 (scenario A) to 7 (scenario
B). There are of course other complications, stemming from the
possibility that most UV photons may be reabsorbed locally; moreover in
Scenario B the formation of sufficient OB stars might have to await
the build-up of larger systems, with deeper potential wells, in which
stars could form more efficiently.

       The above examples have assumed a `standard' IMF, and there is
actually further uncertainty.  If the Population III IMF were biased
towards low-mass stars, the situation resembles inefficient star
formation in that a large fraction of the baryons (not just the rare
3-$\sigma$ peaks) would have to collapse non-linearly before enough UV
had been generated to ionize the IGM. By the time this happened, a
substantial fraction of the baryons could have condensed into low mass
stars. This population could even contribute to the MACHO lensing
events (see section 6).

\subsection{Detecting `pregalaxies' at very high redshift.}

     What is the chance of detecting the ancient `pregalaxies' that
ionized the IGM at some redshift $ z_i>5$? The detectability of these
early-forming systems, of subgalactic mass, depends which of the two
scenarios in 3.2 (above) is nearer the truth.  If B were correct, the
individual high-z sources would have magnitudes of 31, and would be so
common that there would be about one per square arc second all over
the sky; on the other hand, option A would imply a lower surface
density of brighter (and more readily detectable) sources for the first
UV (Miralda-Escud\'e and Rees (1998), Barkana and Loeb (1999)). There are already some constraints from the Hubble Deep Field,
particularly on the number of `miniquasars' (Haiman, Madau and Loeb 
1999).  Objects down to 31st magnitude could be detected from the
ground by looking at a field behind a cluster where there might be
gravitational-lens magnification, but firm evidence is likely to await
NGST.

      Note that scenarios A and B would have interestingly different
implications for the formation and dispersal of the first heavy
elements. If B were correct, there would be a large number of
locations whence heavy elements could spread into the surrounding
medium; on the other hand, scenario A would lead to a smaller number
of brighter and more widely-spaced sources.

\subsection{The `breakthrough' epoch}

     Quasar spectra tell us that the diffuse IGM is almost fully ionized back
to $z=5$, but we do not know when it  in effect became an
HII region.  The IGM would already be inhomogeneous at the time when
the ionization occurred.  The traditional model of expanding HII
regions that overlap at a well defined epoch when `breakthrough' occurs
(dating back at least to  Arons and McCray (1972)) is
consequently rather unrealistic.  By the time ionization
occurs the gas is so inhomogeneous that half the mass (and far more
than half of the recombinations) is within 10 percent of the
volume. HII regions in the `voids' can overlap (in the sense that the
IGM becomes ionized except for `islands' of high density) before even
half the material has been ionized. Thereafter, the overdense regions
would be `eroded away': Stromgren surfaces encroach into them; the
neutral regions shrink and present a decreasing cross-section; the
mean free path of ionizing photons (and consequently the UV background intensity J)
goes up (Miralda-Escud\'e, Haehnelt and Rees 1999, Gnedin 1999).

    The thermal history of the IGM beyond $z= 5$ is relevant to the
modelling of the absorption spectra of quasars at lower redshifts. The recombination and
cooling timescales are comparable to the cosmological
expansion timescale. Therefore the `texture' and temperature of the
filamentary structure responsible for the lines in the Lyman alpha
`forest' yield fossil evidence of the thermal history at higher
redshifts.

\subsection{Black hole formation and AGNs at high {\boldmath $z$}?} 
      
The observations of high-redshift galaxies tell us that some
structures (albeit perhaps only exceptional ones) must have attained
galactic scales by the epoch $z = 5$. Massive black holes (manifested
as quasars) accumulate in the deep potential wells of these larger
systems.  Quasars may dominate the UV background at $z < 3$: if their
spectra follow a power-law, rather than the typical thermal spectrum
of OB stars, then quasars are probably crucial for the second
ionization of He, even if H was ionized primarily by starlight.  (One
interesting point that somewhat blurs this issue has recently been
made by Tumlinson and Shull (1999). They note that, if the metallicity
were zero, there would be no CNO cycle; high-mass stars therefore need
to contract further before reaching the main sequence, and so have
hotter atmospheres, emitting more photons above the He ionization
edge.)

      At redshifts $z = 10$, no large galaxies may yet have assembled,
but CDM-type models suggest that `subgalaxies' would exist. Would
these have massive holes (perhaps `mini-AGNs') in their centres? This
is interesting for at least two reasons: first, the answer would
determine how many high-energy photons, capable of doubly-ionizing He,
were produced at very high redshifts (Haiman and Loeb 1998); second,
the coalescence of these holes, when their host `subgalaxies' merge to
form large galaxies, would be signalled by pulse-trains of
low-frequency gravitational waves that could be detected by
space-based detectors such as LISA (Haehnelt 1994).

    The accumulation of a  central black hole  may require virialised
systems with large masses and deep potential wells (cf Haehnelt and
Rees 1993, Haehnelt and Kauffmann (1999)); if so, we would naturally expect the UV background at
the highest redshifts to be contributed mainly by stars in
`subgalaxies'. However, this is merely an expectation; it could be,
contrariwise, that black holes readily form even in the first $10^8 \Msun$
CDM condensations (this would be an extreme version of a `flattened'
IMF), Were this the case, the early UV production could be dominated
by black holes. This would imply that the most promising high-z
sources to seek at near-IR wavelengths would be miniquasars, rather
than `subgalaxies'. It would also, of course, weaken the connection
between the ionizing background and the origin of the first heavy
elements.

\subsection{Distinguishing between objects with {\boldmath $z> z_i$} and {\boldmath $z< z_i$}}

The blanketing effect due to the Lyman alpha forest -- known to be
     becoming denser towards higher redshifts, and likely therefore to
     be even thicker beyond $z=5$ -- would be severe, and would block
     out the blue wing of Lyman alpha emission from a high-$z$ source.
     Such objects may still be best detected via their Lyman alpha
     emission even though the absorption cuts the equivalent width by
     half. But at redshifts larger than $z_i$ -- in other words, before
     ionization breakthrough -- the Gunn-Peterson optical depth is so
     large that any Lyman alpha emision line is blanketed completely,
     because the damping wing due to IGM absorption spills over into
     the red wing (Miralda-Escud\'e and Rees 1998). This means that any objects detectable beyond $z$
     would be characterised  by a discontinuity at the redshifted Lyman alpha
     frequency. The Lyman alpha line itself would not be detectable
     (even though this may be the most prominent feature in objects
     with $z<z_i$).

\section{RADIO AND MICROWAVE PROBES OF THE IONIZATION EPOCH}

\subsection{CMB fluctuations as a probe of the ionization epoch}

    If the intergalactic medium were suddenly reionized at a redshift
$z$, then the optical depth to electron scattering would be
$\sim 0.02h^{-1} \left((1+z)/10\right)^{3/2}
\left(\Omega_b h^2/0.02\right)$ 
(generalisation to more realistic scenarios of gradual reionization is
straightforward). Even when this optical depth is far below unity, the
ionized gas constitutes a `fog'-- a partially opaque `screen' -- that
attenuates the fluctuations imprinted at the recombination era; the
fraction of photons that are scattered at $z_i$ then manifest a
different pattern of fluctuations, characteristically on larger
angular scales. This optical depth is consequently one of the
parameters that can in principle be determined from CMB anisotropy
measurements (Zaldarriaga, Spergel and Seljak 1997). It is feasible to detect a value as small as 0.1 --
polarization measurements may allow even greater precision, since the
scattered component would imprint polarization on angular scales of a
few degrees, which would be absent from the Sachs-Wolfe fluctuations
on that angular scale originating at $t_{rec}$.

  There are two effects that could introduce secondary fluctuations on
small angular scales. First, the ionization may be patchy on a large
enough scale for irregularities in the `screen' to imprint extra
angular structure on the radiation shining through from the `last
scattering surface at the recombination epoch'. Second, the
fluctuations may have large enough amplitudes for second-order effects
to induce perturbations. (Hu, 1999)

\subsection{21 cm emission,  absorption and tomography}

   The 21 cm line of HI at redshift $z$ would contribute to the
background spectrum at a wavelength of $21(1 + z)$ cm. This
contribution depends on the spin temperature $T_s$ and the CMB
temperature $T_{bb}$. It amounts to a brightness temperature of only
$0.01 h^{-1} (\Omega_b h^2  /0.02)  ((1+z)/10)^{1/2}(T_s
-T_{bb})/T_s \, {\rm K}$ -- very small compared with the 2.7K of the present CMB;
and even smaller compared to the galactic synchrotron radiation that
swamps the CMB, even at high galactic latitudes, at the long
wavelengths where high-$z$ HI should show up.

   Nonetheless, inhomogeneities in the HI may be detectable because
they would give rise not only to angular fluctuations but also to
spectral structure. 
(Madau, Meiksin and Rees 1997, Tozzi \etal 1999) 
If the same strip of sky were scanned at two radio
frequencies differing by (say) 1 MHz, the temperature fluctuations due
to the CMB itself, to galactic thermal and synchrotron backgrounds,
and to discrete sources would track each other closely.  Contrariwise,
there would be no correlation between the 21 cm contributions, because
the two frequencies would be probing `shells' in redshift space whose
radial separation would exceed the correlation length.  It may
consequently be feasible to distinguish the 21 cm background,
utilizing a radio telescope with large collecting area. The fact that
line radiation allows 3-dimensional tomography of the high-$z$ HI
renders this a specially interesting technique.

     For the 21 cm contribution to be observable, the spin temperature
$T_s$ must of course differ from $T_{bb}$.  The HI would be detected
in absorption or in emission depending on whether $T_s$ is lower or
higher than $T_{bb}$.  During the `dark age' the hyperfine levels of
HI are affected by the microwave background itself, and also by
collisional processes.  $T_s$ will therefore be a weighted mean of the
CMB and gas temperatures.  Since the diffuse gas is then cooler than
the radiation (having expanded adiabatically since it decoupled from
the radiation), collisions  would tend to lower $T_s$ below $T_{bb}$ , so that the 21 cm line
would appear as an absorption feature, even in the CMB. At the low densities of the IGM, collisions are however ineffectual in lowering $T_s$ substantially below $T_{bb}$ (Scott and Rees 1990).   When the
first UV sources turn on, Lyman alpha (whose profile is itself controlled
by the kinetic temperature) provides a more effective coupling between
the spin temperature and the kinetic temperature.  If Lyman alpha
radiation penetrates the HI without heating it, it can actually lower
the spin temperature so that the 21 cm line becomes a stronger
absorption feature.  However, whatever objects generate the Lyman
alpha emission would also provide a heat input, which would soon raise
$T_s$ above $T_{bb}$.

       When the kinetic temperature rises above $T_{bb}$, the 21 cm
feature appears in emission.  The kinetic temperature can rise due to
the weak shocking and adiabatic compression that accompanies the
emergence of the first (very small scale) non-linear structure (cf
section 2). When photoionization starts, there will also, around each
HII domain, be a zone of predominantly neutral hydrogen that has been
heated by hard UV or X-ray photons (Tozzi \etal (1999). This latter
heat input  would be more important if the first UV sources emitted
radiation with a power-law (rather than just exponential) component.

 In principle, one might be able to detect incipient large-scale
structure, even when still in the linear regime, because it leads to
variations in the column density of HI, per unit redshift interval,
along different lines of sight (Scott and Rees (1990)).

     Because the signal is so weak, there is little prospect of
detecting high-$z$ 21 cm emission unless it displays structure on
(comoving) scales of several Mpc (corresponding to angular scales of
several arc minutes) According to CDM-type models, the gas is likely
to have been already ionized, predominantly by numerous ionizing
sources each of sub-galactic scale, before such large structures
become conspicuous. On the other hand, if the primordial gas were
heated by widely-spaced quasar-level sources, each of these would be
surrounded by a shell that could feasibly be revealed by 21cm
tomography using, for instance, the new Giant Meter Wave Telescope
(GMRT) (Swarup (1994)).  With luck, effects of this kind may be
detectable. Otherwise, they will have to await next-generation
instruments such as the Square-Kilometer Array.

\section{VERY DISTANT SUPERNOVAE\\ (AND PERHAPS GAMMA-RAY BURSTS)}

\subsection{The supernova rate at high redshifts}

         If the reheating and ionization were due to OB stars, it is
straightforward to calculate how many supernovae would have gone off,
in each comoving volume, as a direct consequence of this output of UV,
also how many supernovae would be implicated in producing the heavy elements 
detected in quasar absorption lines: there would be one, or maybe several, per year in
each square arc minute of sky (Miralda-Escud\'e and Rees 1997).  The
precise number  depends partly on the redshift and the cosmological model,
but also on the uncertainties about the UV background, and about the
actual high-$z$ abundance of heavy elements. 

   These high-$z$ supernovae would be primarily of Type 2.  The typical
observed light curve has a flat maximum lasting 80 days. One would
therefore (taking the time dilation into account) expect each
supernova to be near its maximum for nearly a year. It is possible
that the explosions proceed differently when the stellar envelope is
essentially metal-free, yielding different light curves, so any
estimates of detectability are tentative. However, taking a standard
Type 2 light curve (which may of course be pessimistic), one
calculates that these objects should be approximately 27th magnitude in
J and K bands even out beyond $z = 5$.  The detection of such objects
would be an easy task with the NGST (Stockman 1998). With existing facilities it is
marginal. The best hope would be that observations of clusters of
galaxies might serendipitously reveal a magnified
gravitationally-lensed image from far behind the cluster.

      The first supernovae may be important for another reason: they
may generate the first cosmic magnetic fields. Mass loss (via winds or
supernovae permeated by magnetic flux) would disperse magnetic flux
along with the heavy elements. The ubiquity of heavy elements in the
Lyman alpha forest indicates that there has been widespread diffusion
from the sites of these early supernovae, and the magnetic flux could
have diffused in the same way. This flux, stretched and sheared by
bulk motions, can be the `seed' for the later amplification processes
that generate the larger-scale fields pervading disc galaxies.

\subsection{Gamma ray bursts: the most luminous known cosmic objects }

   Some subset of massive stars may give rise to gamma-ray
bursts.
It may indeed turn out that all the long-duration bursts detected by Beppo-SAX involve some supernova-type event, and that the shorter bursts (maybe less highly beamed) are caused by compact binary coalescence at more modest redshifts.
 Bursts have already been detected out to $z=3.4$; their optical
afterglows are 100 times brighter than supernovae. Prompt optical
emission concurrent with the 10-100 seconds of the burst itself
(observed in one case so far, but expected in others) is more luminous
by a further factor 100. Gamma-ray bursts are, however, far rarer than
supernovae -- even though the afterglow rate could exceed that of the
bursts themselves if the gamma rays were more narrowly beamed than the
slower-moving ejecta that cause the afterglow.  Detection of
ultra-luminous optical emission from bursts beyond $z=5$ would offer a
marvellous opportunity to obtain a high-resolution spectrum of
intervening absorption features. (Lamb and Reichart 1999)
 
\section{WHERE ARE THE OLDEST (AND THE\\ EXTREME METAL-POOR)  STARS?}

   The efficiency of early mixing is important for the interpretation
of stars in our own galaxy that have ultra-low metallicity -- lower
than the mean metallicity of $10^{-2}-10^{-3}$ times solar that is likely to have been generated in
association with the UV background at $z > 5$.  
For a comprehensive review of what is known about such stars, see Beers (1999).
If the heavy elements
were efficiently mixed, then these stars would themselves need to have
formed before galaxies were assembled.   The mixing, however, is
unlikely to operate on scales as large as a protogalaxy -- if it did,
the requisite bulk flow speeds would be so large that they would
completely change the way in which galaxies assembled, and would certainly need to be incorporated in simulations of the Lyman alpha forest.

   As White and Springel (1999) have recently emphasised, it is
important to distinguish between the first stars and the most
metal-poor stars. The former would form in high-sigma peaks that would
be correlated owing to biasing, and which would preferentially lie
within overdensities on galactic scales. These stars would therefore
now be found within galactic bulges. However, most of the metal-poor
stars could form later. They would now be in the halos, of galaxies,
though they would not have such an extended distribution as the dark
matter. This is because they would form in subgalaxies that would
tend, during the subsequent mergers, to sink via dynamical friction
towards the centres of the merged systems. There would nevertheless be
a correlation between metallicity, age and kinematics within the
Galactic Halo.  This is a project where NGST could be crucial,
especially if it allowed detection of halo stars in other nearby
galaxies.

       The number of such stars depends on the IMF. If this were flat,
there would be fewer low-mass stars formed concurrently with those
that produced the UV background. If, on the other hand, the IMF were
initially steep, there could in principle be a lot of very low mass
(MACHO) objects produced at high redshift, many of which would end up
in the halos of galaxies like our own.

\section{SUMMARY}
   
    Perhaps only 5 percent of star formation occurred before $z=5$
(the proportion could be higher if most of their light were
reprocessed by dust). But these early stars were important: they
generated the first UV and the first heavy elements; they provided the
backdrop for the later formation of big galaxies and larger-scale
structure.  Large-scale structure may be elucidated within the next
decade, by ambitious surveys (2-degree field and Sloan) and studies of
CMB anisotropies; as will be the evolution of galaxies and their
morphology.  The later talks in this conference will highlight the
exciting progress and prospects in this subject.  But despite this
progress, we shall, for a long time, confront uncertainty about
the efficiency and modes of star formation in early structures on
subgalactic scale.

I am grateful to my collaborators, especially Tom Abel, Zoltan Haiman,
Martin Haehnelt,  Avi Loeb, Jordi Miralda-Escud\'e, Piero Madau, Avery
Meiksin, and Paulo Tozzi, for discussion of the topics described here.
I am also grateful to the Royal Society for support.


\begin{thebibliography}{}
\bibitem{} Abel, T., Bryan, G.L. \& Norman, M. L., 1999 ApJ, in press
\bibitem{} Arons, J. \& Wingert, D. W.   1972 ApJ, 177, 1
\bibitem{} Barkana, R. \& Loeb, A. 1999 ApJ, in press
\bibitem{} Beers, T.C. 1999 in `The First Stars' (ESO publications)
\bibitem{} Bromm, V., Coppi, P.S. \& Larson, R.B. 1999 ApJ, in press
\bibitem{} Ciardi B., \& Ferrara, A., 1997 ApJ, 483, L5
\bibitem{}Ciardi, B., Ferrara, A., \& Abel, T., 1999 ApJ, in  press.
\bibitem{} Gnedin, N. 1999 ApJ, in press 
\bibitem{} Gnedin, N. \& Ostriker, J.P. 1998  ApJ,  486, 581
\bibitem{} Haiman, Z., Loeb, A. 1998 ApJ, 503, 505
\bibitem{} Haiman, Z., Rees, M. J. \& Loeb, A.   1997 ApJ, 476, 458 
\bibitem{} Haehnelt, M. 1994 MNRAS, 269, 199
\bibitem{} Haehnelt, M. and Kauffmann, G. 1999 MNRAS, in press
\bibitem{} Haehnelt, M. \& Rees, M.J. 1993 MNRAS, 263, 168
\bibitem{} Haiman, Z., Abel, T \& Rees, M.J. 1999 ApJ, in press
\bibitem{} Haiman, Z. \& Loeb, A., 1998 ApJ, 503, 505
\bibitem{}  Haiman, Z., Rees, M.J.  \& Loeb, A., 1996 ApJ, 467, 522 
\bibitem{}  Haiman, Z., Madau, P. \& Loeb, A.  1999 ApJ, 514, 535 
\bibitem{} Hu, W. 1999 ApJ, in press
\bibitem{} Lamb, D. Q. \& Reichart, D.E. 1999 ApJ, in press.
\bibitem{} Loeb, A., 1999  in proceedings of conference in honour of H. Spinrad (Astr. Soc.
Pacific) in press
\bibitem{} Madau, P., Hardt, F. \& Rees,M.J. 1999 ApJ, 514, 648
\bibitem{} Madau, P., Meiksin, A. \& Rees, M.J. 1997 ApJ, 475, 429
\bibitem{} Miralda-Escud\'e, J.,  Haehnelt, M \& Rees, M. J, 1999 ApJ, in press.
\bibitem{} Miralda-Escud\'e, J. \& Rees, M. J., 1997 ApJ (letters) 478, L57
\bibitem{} Miralda-Escud\'e, J. \& Rees, M. J., 1998 ApJ,  497, 21
\bibitem{} Scott, D. \& Rees, M.J. 1990 MNRAS, 247, 510.
\bibitem{}  Stecher, T.P.  \& Williams, D.A., 1967 ApJ (Letters)  149, L1 
\bibitem{} Stockman, P. 1998. Proceedings of first NGST conference (NASA publications)
\bibitem{} Swarup, G. 1994 `The GMRT'. TIFR report
\bibitem{} Tegmark, M., Silk, J., Rees, M. J., Blanchard, A., Abel, T.
\& Palla, F.  1997  ApJ, 474, 1
\bibitem{} Tozzi, P., Madau, P., Meiksin, A. \& Rees, M.J. 1999 ApJ, in press
\bibitem{} Tumlinson, J. and Shuff, J.M. 1999 ApJ, in press
\bibitem{} White, S.D.M. \& Springel, V. 1999 in `The First Stars' (ESO publications, Munich)
\bibitem{} Zaldarriaga, M., Spergel, D.N. \& Seljak U. 1997 ApJ, 488, 1.
\end{thebibliography}
\end{document}